\documentclass[a4paper,11pt]{article}
\usepackage{float}
\usepackage{jinstpub} 
\usepackage{lineno}
\usepackage{graphicx}
\usepackage{subfigure}
\usepackage{caption}
\usepackage{subcaption}
\usepackage{comment}
\usepackage{appendix}
\usepackage{adjustbox}

\title{\boldmath Fast Low Energy Reconstruction using Convolutional Neural Networks}

\usepackage{jinstpub}
\usepackage[T5,T1]{fontenc}

\author[15]{R. Abbasi,}
\author[63]{M. Ackermann,}
\author[16]{J. Adams,}
\author[38,a]{S. K. Agarwalla,}
\author[9]{J. A. Aguilar,}
\author[20]{M. Ahlers,}
\author[21]{J.M. Alameddine,}
\author[42]{N. M. Amin,}
\author[40]{K. Andeen,}
\author[12]{C. Arg{\"u}elles,}
% ...
\affiliation[9]{Universit{\'e} Libre de Bruxelles, Science Faculty CP230, B-1050 Brussels, Belgium}
% ...
\affiliation[12]{Department of Physics and Laboratory for Particle Physics and Cosmology, Harvard University, Cambridge, MA 02138, USA}
% ...
\affiliation[15]{Department of Physics, Loyola University Chicago, Chicago, IL 60660, USA}
\affiliation[16]{Dept. of Physics and Astronomy, University of Canterbury, Private Bag 4800, Christchurch, New Zealand}
% ...
\affiliation[20]{Niels Bohr Institute, University of Copenhagen, DK-2100 Copenhagen, Denmark}
\affiliation[21]{Dept. of Physics, TU Dortmund University, D-44221 Dortmund, Germany}
% ...
\affiliation[38]{Dept. of Physics and Wisconsin IceCube Particle Astrophysics Center, University of Wisconsin{\textemdash}Madison, Madison, WI 53706, USA}
% ...
\affiliation[40]{Department of Physics, Marquette University, Milwaukee, WI 53201, USA}
% ...
\affiliation[42]{Bartol Research Institute and Dept. of Physics and Astronomy, University of Delaware, Newark, DE 19716, USA}
% ...
\affiliation[63]{Deutsches Elektronen-Synchrotron DESY, Platanenallee 6, D-15738 Zeuthen, Germany}
\affiliation[a]{also at Institute of Physics, Sachivalaya Marg, Sainik School Post, Bhubaneswar 751005, India}
% ...

\emailAdd{analysis@icecube.wisc.edu}

\abstract{IceCube is a Cherenkov detector instrumenting over a cubic kilometer of glacial ice deep under the surface of the South Pole. The DeepCore sub-detector lowers the detection energy threshold to a few GeV, enabling the precise measurements of neutrino oscillation parameters with atmospheric neutrinos. The reconstruction of neutrino interactions inside the detector is essential in studying neutrino oscillations. It is particularly challenging to reconstruct sub-100 GeV events with the IceCube detectors due to the relatively sparse detection units and detection medium. 
Convolutional neural networks (CNNs) are broadly used in physics experiments for both classification and regression purposes. This paper discusses the CNNs developed and employed for the latest IceCube-DeepCore oscillation measurements~\cite{paper_flercnn}. These CNNs estimate various properties of the detected neutrinos, such as their energy, direction of arrival, interaction vertex position, flavor-related signature, and are also used for background classification.

}


\begin{document}
\maketitle
\flushbottom

\section{Introduction}\label{sec:intro}

\begin{figure}[tbh]
\centering
{\includegraphics[width=0.8\columnwidth]{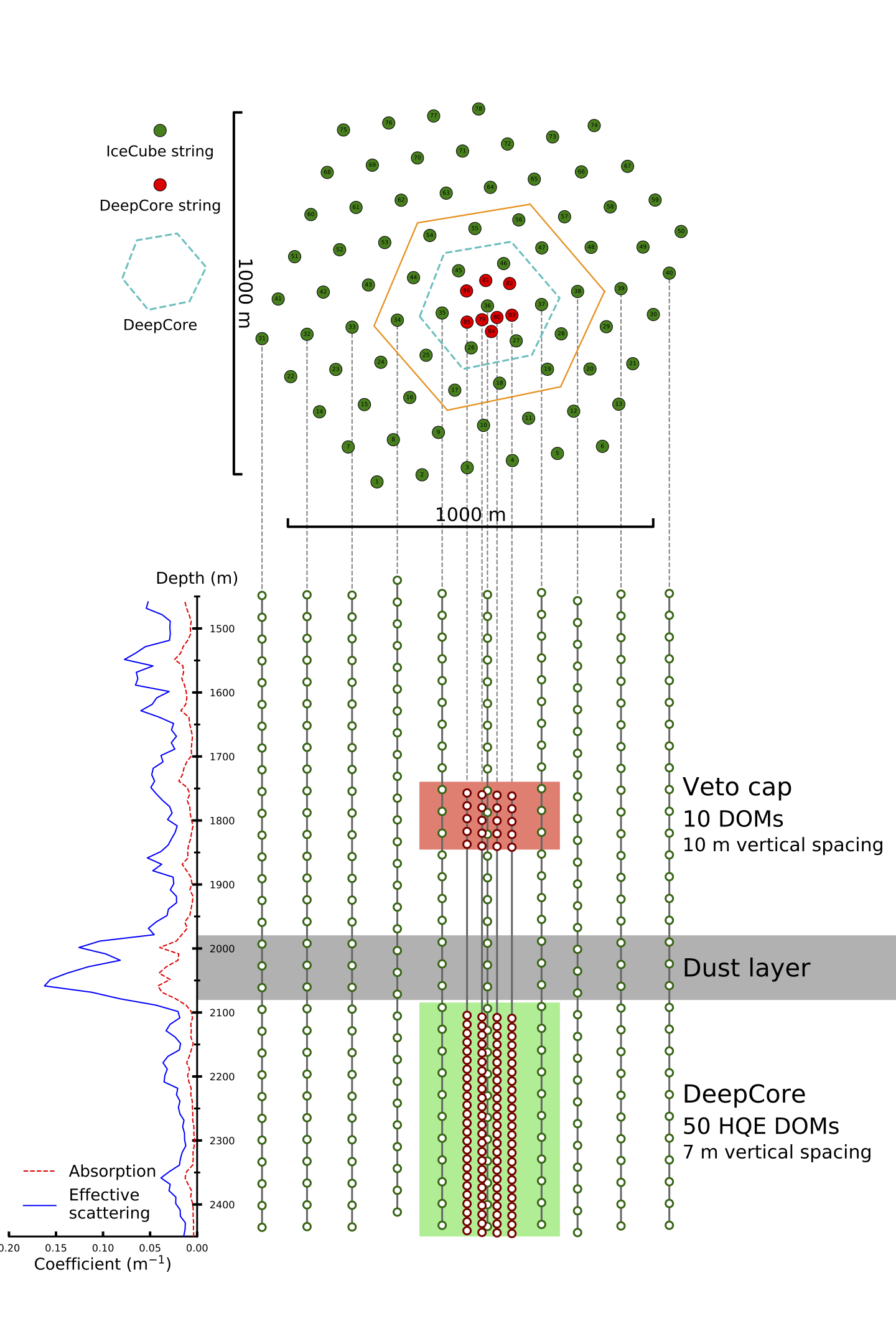}}
\caption{
The upper panel shows a top view of the IceCube detector. DeepCore strings are indicated by red circles and IceCube strings by green circles. The DeepCore detector consists of both DeepCore strings and IceCube strings located within the dashed green hexagonal circle. In this work, the CNNs use only the strings within the orange hexagon. The lower panel shows a side view of the detector. The green shading highlights the DeepCore DOMs and the red shading highlights the veto DOMs.}
\label{fig:detector}
\end{figure}

The IceCube Neutrino Observatory at the South Pole, consists of 5,160 digital optical modules (DOMs) instrumenting over one cubic kilometer of glacial ice. 
Each DOM consists of a photomultiplier tube (PMT) oriented downward to detect Cherenkov photons, along with electronic components that support and maintain the operation of the PMT. These components are housed and protected by a clear spherical glass shell~\cite{Aartsen_2017}. 
A group of 60 DOMs, connected vertically by a cable, makes one string.
The main IceCube array consists of 78 strings and has detected neutrinos up to PeV-energies. The DeepCore sub-detector, in the bottom center of the main array, is made of 8 strings, which have a denser spatial configuration and use higher quantum efficiency PMTs. 
With the instrumentation of DeepCore, it is possible to measure and reconstruct neutrino interactions at GeV energies, providing excellent sensitivity to atmospheric neutrino oscillations and in particular to atmospheric $\nu_\mu$ disappearance~\cite{paper_flercnn}. The top and side views of IceCube and DeepCore are shown in Figure~\ref{fig:detector}. 

When neutrinos interact in or near the IceCube detector and produce charged particles, the relativistic charged particles propagate in the ice and give rise to Cherenkov photons. DOMs record data when the PMT signal voltage passes a threshold equivalent to one quarter the amplitude of an average photoelectron (PE) response~\cite{10.1063/1.4902795}. 
To filter out random noise, a local-coincidence trigger is applied, requiring that signals from multiple DOMs occur within a predefined time window~\cite{paper_verification}.
Once a trigger is formed, the PMT signals are processed by the analog to digital converters into digitized waveforms. Waveforms are converted into pulses using a decomposition into a basis of templates of the known response of the PMT to a photoelectron. Using the charge and time information extracted from the pulses of each DOM, algorithms~\cite{Abbasi_2021,gnn,le_reco_paper} are employed to reconstruct the neutrino interactions in the detector.

Convolutional neural networks (CNN) are deep learning models that excel in processing images. By learning features through convolutional layers, CNNs can efficiently extract patterns from complex data. In reconstruction applications for particle interactions in large detectors~\cite{Aurisano_2016,Baldi_2019,Abbasi_2021}, CNNs are increasingly employed to interpret high-dimensional data and enhance accuracy in tasks such as event classification and signal processing.
In previous IceCube reconstruction studies for high-energy astrophysical neutrinos, a CNN was developed to process data from all 86 strings~\cite{Abbasi_2021}. 
In this work, we introduce a different approach that uses CNNs specifically optimized to reconstruct sub-100 GeV neutrino interactions concentrated in the DeepCore detector.
The results of this low-energy CNN were employed in a recent atmospheric neutrino oscillation measurement~\cite{paper_flercnn}, and the aim of this paper is to provide the further technical details that complement the published results.

\section{Input data and Neural Network Architecture}\label{sec:cnn}
\begin{figure}[tb]
\center
\begin{minipage}{.6\textwidth}
\center{\includegraphics[width=\columnwidth]{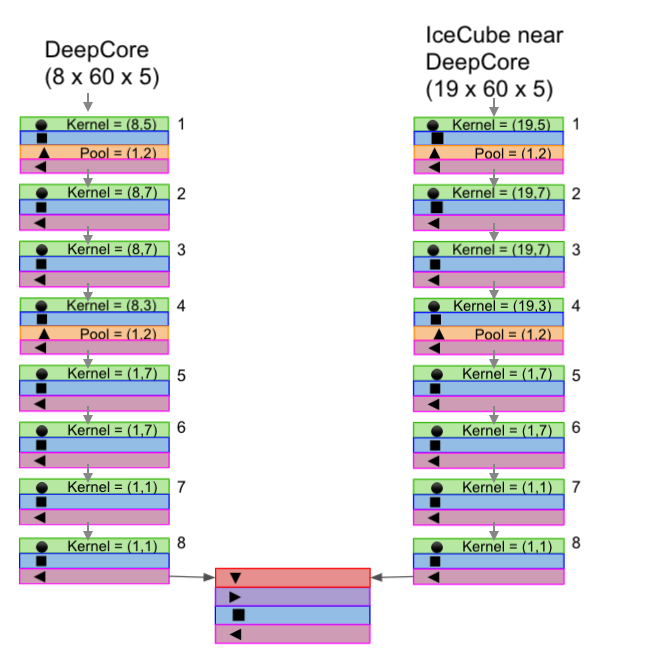}}
\end{minipage}\hspace{4em}%
\begin{minipage}{.2\textwidth}
 \center{\includegraphics[width=\columnwidth]{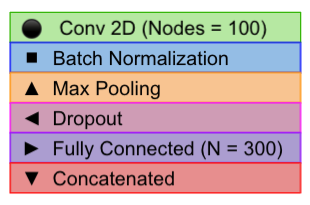}}
\end{minipage}
\caption{\label{fig:Architecture}Two-branch architecture of the CNN, including a legend for names of layers on the right. Eight DeepCore and 19 IceCube strings with 60 DOMs on each string and 5 summarized variables from each DOM from the digitized pulses are fed into the two sub-networks correspondingly. The size of the kernel is listed in the graph and spans only along the $z$-depth axis.}
\end{figure}

The CNNs in this study are employed for the five distinct use cases to reconstruct low-energy neutrino interactions: the estimation of the neutrino's energy, arrival direction, and interaction vertex position, and the binary classification between track-like signal events, $\nu_{\mu}$ charged-current (CC) interactions versus cascade-like background events, including $\nu_{\mu}$ neutral-current (NC) interactions, and background atmospheric muon events versus signal neutrino events.
The same CNN architecture, with different output layers for different training goals, was trained for five distinct use cases, separately reconstructing different physical properties of a particle interaction in the detector. Another advantage of training five CNNs instead of a combined one is to optimize the performance of each training by employing specialized training samples to ensure a balanced distribution of events in the training sets, as discussed in Section~\ref{sec:train}. 
The shared architecture and the necessary difference in the output layers are discussed in detail in this section. 
For GeV-scale neutrino reconstruction in DeepCore, it is most productive to use data from in and immediately around the DeepCore region, rather than the full detector. The CNNs use all eight DeepCore strings and the 19 surrounding IceCube strings, which form the two rings outside the DeepCore region, as shown in the orange hexagon in Figure~\ref{fig:detector}. These two rings help optimize reconstruction performance in the DeepCore region by providing additional information from adjacent IceCube DOMs, while avoiding the computational cost of including farther DOMs, which typically do not provide meaningful data for reconstructing neutrino interactions within DeepCore. Due to differences in spatial distance and quantum efficiency between DOMs on the DeepCore and IceCube strings, data from these two categories are fed into separate input layers, each with a dimension of [N$_{\rm{string}}$, 60, 5], where N$_{\rm{string}}$ represents the number of strings from which data is included (19 and 8 for the IceCube and DeepCore input layers respectively) and 60 corresponds to the 60 DOMs on each string. When the datasets are processed, the input is treated as an image of shape (N$_{\rm{string}}$ × 60) with 5 channels representing the summarized variables (see below) per DOM. 
The 1D CNN kernel slides along the vertical, or $z$-depth, of the strings with the kernel size given in Figure~\ref{fig:Architecture}, ranging from taking a maximum of 2 DOMs above and below the target DOM in the convolution. Since the DeepCore strings are deployed in an irregular array, no additional convolution is applied in the $xy$-plane. This is a simplified architecture based on~\cite{Aurisano_2016}, which performs full convolutions over the hexagonally distributed IceCube strings. As an initial attempt to apply CNNs for reconstructing low-energy events in DeepCore, this work demonstrates competitive performance. More recent applications using Graph Neural Networks~\cite{gnn} offer greater generality and are adopted in future analyses. The same convolutional kernel is applied to the IceCube strings to be consistent, and thus both branches of the CNN use the same kernel applied in $z$-depth only. Technically, the 1-D convolution is realized by a ``Conv 2D" layer.

After passing through functional layers, such as ``Conv 2D", ``Batch Normalization", ``Max Pooling", and ``Dropout" in the sub-networks (see Figure~\ref{fig:Architecture} for details), the outputs of the two sub-networks are concatenated and fed into one fully connected dense layer. This last layer is then used to estimate the physical properties, which serve as the labels for the CNN.

The data from each DOM is summarized from the charge and time information of the digitized pulses recorded in the time window of the event interaction, 
typically within -500 to 4000 ns of the DeepCore trigger time~\cite{IceCube:2011ucd}. This time window corresponds to approximately eight times the scattering length divided by the speed of light in ice, so pulses falling outside this window are generally considered noise. To ensure consistency across all data and simulations, we quantize the charges in steps of 0.05 PE before using them to calculate the CNN input features. Five summary variables are calculated from the pulse series for each DOM to represent the statistical features of the digitized waveform: total charge, time of the first pulse, time of the last pulse, charge weighted mean of pulse times, and charge weighted standard deviation of pulse times. 

These summary variables have significantly different characteristic scales. To train the networks more efficiently, we rescale the input variables by factors, which are listed in Table~\ref{tab:input_features}. These are chosen such that the variables remain near the range [-1,1].

\begin{table}[htbp]
\centering

\caption{\label{tab:input_features} Scaling factors to transform input features to values near the range [-1, 1].}
\begin{tabular}{|c | c | c |}
 \hline
 \textbf{Input Variable} & \textbf{ \begin{tabular}{@{}c@{}} Scaling factor\end{tabular}} \\[0.5ex] 
 \hline
Sum Charge & 1/25  \\
\hline
Time First Pulse & 1/4000  \\
\hline
Time Last Pulse & 1/4000  \\
\hline
Charge Weighted Mean & 1/4000 \\
\hline
\begin{tabular}{@{}c@{}}Charge Weighted\\ Standard Deviation\end{tabular} & 1/2000 \\
\hline
\end{tabular} \\
\end{table}

In the output layer, the CNNs use linear activation functions for estimating arrival direction (zenith angle) and interaction vertex coordinates ($x$, $y$, $z$), and rectified linear units (ReLU) for predicting neutrino energy.
The difference is because zenith angle, in units of radians, and vertex coordinates, in units of meters, do not have hard boundaries while energy must be positive. For the binary classifications, a binary entropy loss function was used. The two classifiers have different classification tasks and therefore employ different training samples, see Sec.~\ref{sec:train} for details, and cannot be combined into a multi-classifier.

\section{Loss function for CNN training}\label{app:loss}

\begin{table}[!bp]
\centering
\caption{\label{tab:lossfunctions} Loss functions used for each of the networks.}
\smallskip
\begin{tabular}{|c | c |} 
 \hline
 CNN reconstruction & Loss function\\[0.5ex] 
 \hline
 Energy & Mean Absolute Percentage Error \\
 \hline
Zenith &  Mean Squared Error \\
\hline
Vertex & Mean Squared Error\\
\hline
Track & Binary Cross Entropy \\
\hline
Muon & Binary Cross Entropy \\
\hline
\end{tabular} \\
\end{table}

The loss functions are specially designed to optimize the reconstruction performance for different physics variables. Table~\ref{tab:lossfunctions} has the summary of the losses for all the CNNs. The energy CNN uses the mean absolute percentage error loss, such that
\begin{equation}
\textrm{Loss} = \frac{100\%}{n}\sum_{i=1}^{n}\left |\frac{t_i-r_i}{t_i}\right|
\end{equation}
where $n$ is the number of events per batch, $t_i$ is the true value and $r_i$ is the reconstructed (or estimated) variable by the CNN. This was chosen carefully such that the energy network would give equal importance to reconstructing the low-energy events (at a few GeV) as the high-energy events (at 100 GeV). Since the $\mathcal{O}$(10) GeV events are the region where oscillation is expected, the percentage error ensures that the 10 GeV events are well-reconstructed without over-emphasis on the $\mathcal{O}$(100) GeV events. The vertex and zenith networks use the mean squared error loss:
\begin{equation}
\textrm{Loss} = \frac{1}{n}\sum_{i=1}^{n}\left ({t_i-r_i}\right)^2.
\end{equation}
The CNNs for classification purposes use a binary cross-entropy loss function:
\begin{equation}
   \textrm{Loss} = -\frac{1}{n}\sum_{i=1}^{n}{(t_i\ln(r_i) + (1 - t_i)\ln(1 - r_i))},
\end{equation}
where $t_i$ is the MC truth label for event $i$ (1 for signal, 0 for background), and $r_i$ is the predicted probability of being a signal event.

\section{Training Samples}\label{sec:train}

Simulated Monte-Carlo (MC) datasets of neutrino interactions and atmospheric muons were used to train the CNNs. In the pre-processing, data quality cuts and random shuffling were applied to ensure the desired distributions were used to train the different CNNs, as discussed further below.

The neutrino MC datasets used in this study for training were generated with the GENIE package (version 2.12.8)~\cite{GENIE}. This dataset has larger statistics and is independent of the MC events that were used in the neutrino oscillation data analyses~\cite{paper_flercnn}. The final level analysis cuts based on the outcomes of this study are relatively tight compared to the ones that are used as the quality cuts for the training samples. The muon MC samples were generated with MuonGun, an IceCube software package that simulates atmospheric muons from cosmic rays (method based on~\cite{BECHERINI20061}). The simulated muon events used for training were from a subset of what is used in the oscillation analysis. 
IceCube neutrino analyses typically involve filters, applied successively, to remove background-like events and keep high-quality neutrino candidates. 
The training sets for the CNNs were processed through the same sets of filters and basic reconstruction routines commonly used for  DeepCore oscillation data analysis (see Section III in~\cite{paper_verification} for details).
However, a boosted decision tree (BDT)~\cite{bdt} was trained using lower-level variables and applied to remove neutrino-like background events, mostly atmospheric muons, therefore noted as BDT$_\mu$ (see Section III.C of ~\cite{paper_verification} for details).  
To drop the noise-like events and keep the track-like $\nu_\mu$ charged-current (CC) events, the score of the BDT$_\mu$ was required to be at least 0.95 in all of the CNN training datasets, except for energy, zenith, and the atmospheric muon dataset used for training the atmospheric muon classification. 
To retain enough statistics, the atmospheric muon events used for training were taken from the nominal dataset before the BDT$_\mu$ filter was applied.

The detailed composition, statistics, and energy range of each training dataset are summarized in Table~\ref{tab:trainingsamples}. The differences in quality cuts that were applied during the training sets' preparation for the 5 CNNs for optimized performance are summarized in Table~\ref{tab:uniquecuts}. The details of the training sets are explained and described as the following. 
\begin{table}[htbp]
\centering
\caption{\label{tab:trainingsamples} Composition of datasets and energy ranges of events used for training each network.}
\begin{tabular}{|c | c | c | c |} 
 \hline
 \textbf{CNN task} & \textbf{Particle} & \textbf{\# Training Events} & \textbf{Energy Range (GeV)}  \\[0.5ex] 
 \hline
 Energy & $\nu_\mu$ CC & 9 million & 1-500 \\
 \hline
Zenith  & $\nu_\mu$ CC & 5 million & 1-300 \\
\hline
Vertex & $\nu_\mu$ CC & 5 million & 1-500 \\
\hline
Muon & $\nu_\mu$, $\nu_e$ (CC/NC), $\mu$ & 7 million & $\nu$: 5-200, $\mu$: 150-5000  \\
\hline
Track & $\nu_\mu$, $\nu_e$ (CC/NC) & 5 million & 5-200  \\
\hline
\end{tabular} \\
\end{table}
\begin{table}[htbp]
\centering
\caption{\label{tab:uniquecuts} Cuts applied in each training and testing sets. }
\begin{tabular}{|c | c | c | c | c |} 
 \hline
 \textbf{CNN} & $N_{\rm{pulse}} >=$ 8 &  BDT$_\mu>$ 0.95 &other cuts  \\[0.5ex] 
 \hline
Energy &  False & False & $N_{\rm{DOM}} >=$ 7  \\
 \hline
Zenith & False & False & containment; E in [5, 300] GeV \\
 \hline
Vertex & True & True & None \\
 \hline
Track & True & True & None \\
 \hline
Muon & True & True & None \\
\hline
\end{tabular} \\
\end{table}

\subsection{Neutrino Energy}

A sample of over 9 million unweighted $\nu_\mu$ CC events was generated to train the CNN for energy estimation. Since the target energy range for oscillation analyses in DeepCore using atmospheric neutrinos is 5-100 GeV, the training set was simulated over a broader range (1-500 GeV) to ensure sufficient coverage. The goal was to ensure a uniform energy distribution in the training data, giving the network an equal opportunity to encounter events across the entire energy spectrum. Low-level filters and selection criteria, however, removed a disproportionate number of high-energy events. After generating the MC events, we randomly discarded and shuffled them to achieve approximately 20,000 events per 1 GeV energy bin.
During the training, it was found that the events with small number of DOMs ($N_{\rm{DOM}}$) hit were difficult for the network to predict energy. Thus, a cut was introduced to enforce that the event reconstructed by the CNN must have at least 7 DOMs reading out input pulses.

\subsection{Neutrino Zenith Angle}

A dataset of $\nu_\mu$ CC events between 5-300 GeV assuming an injection spectrum that is uniformly distributed in zenith angle in [0, $\pi$] was simulated for the zenith training. To produce a training dataset with a uniform distribution of zenith angles, events were rejected until each 0.01*$\pi$ bin contained a similar number of events. After processing, 5 million simulated events remained for training and testing. 
Training the zenith network, however, relies on showing the network the full track signature. Thus, a cut was applied so that the CNN training set contained only events where the paths of all of the secondary particles, produced in the neutrino interaction, were contained within the active detector volume region indicated by the orange hexagon in Figure~\ref{fig:detector}.
This cut is referred to as ``containment" in Table~\ref{tab:uniquecuts} and corresponds to the variable restrictions ($- 505$ m < $z$ < 500 m and within 260 m to the center-most string, $r_{\rm string36}$).

\subsection{Neutrino Interaction Vertex}
To predict the interaction vertex of a neutrino event, the CNN was trained using a portion of the energy sample, which included approximately 5 million $\nu_\mu$ CC events. 
Since vertex reconstruction was more sensitive to the topology of the deposited photons of the event, additional data-quality cuts of $N_{\rm{pulse}} \geq$ 8 and BDT$_\mu>0.95$ were enforced so that the events for vertex reconstruction training had at least 8 pulses and were neutrino-like.

\subsection{Particle Identification Classifier}

The training dataset for the particle identification (PID) classifier was enforced to have a 50:50 distribution of track-like:cascade-like events by using $\nu_\mu$ CC events for track-like and $\nu_e$ CC and NC events for cascade-like events. This was enforced per energy bin, so that across the energy range (5-200 GeV), there were $~$15,000 track events and $~$15,000 cascade events per 1 GeV. In total, it contained 5 million events.

\begin{figure}[!th]
    \centering
\includegraphics[width=\linewidth]{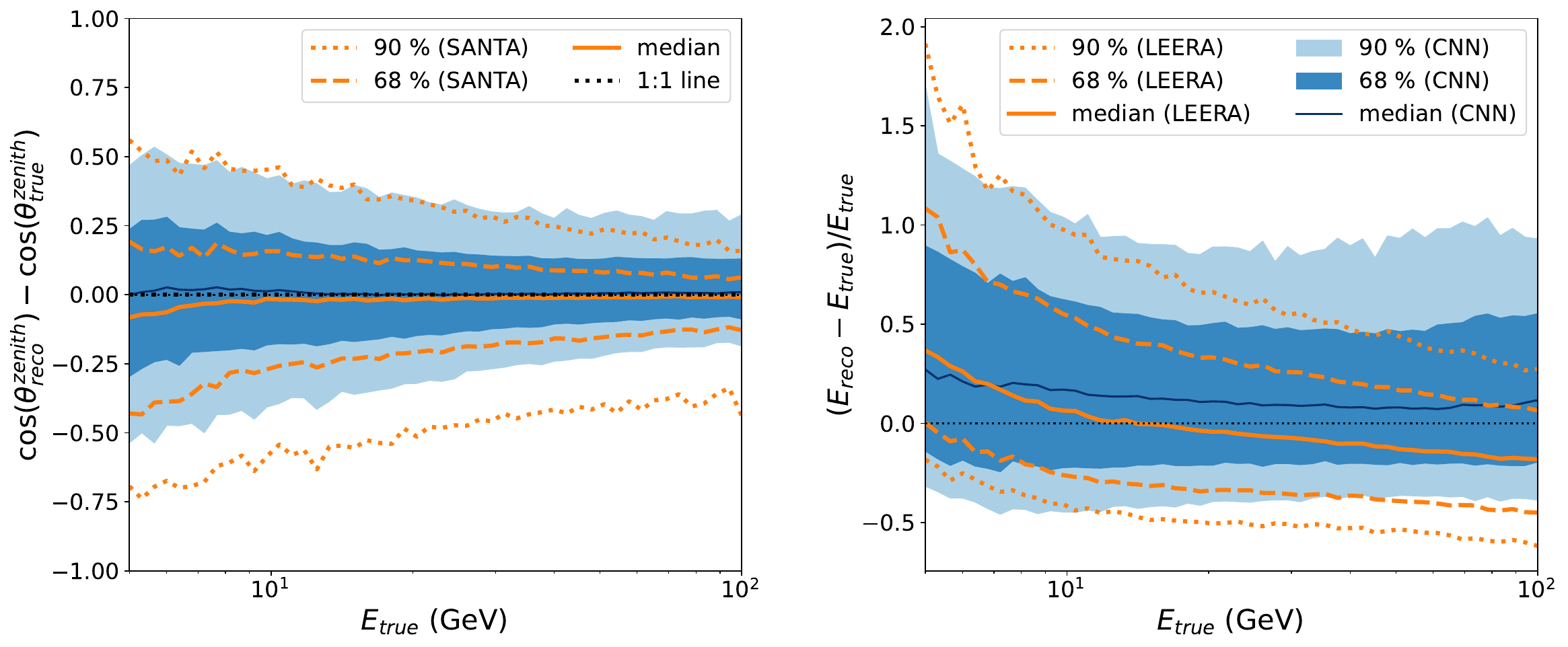}
\includegraphics[width=\linewidth]{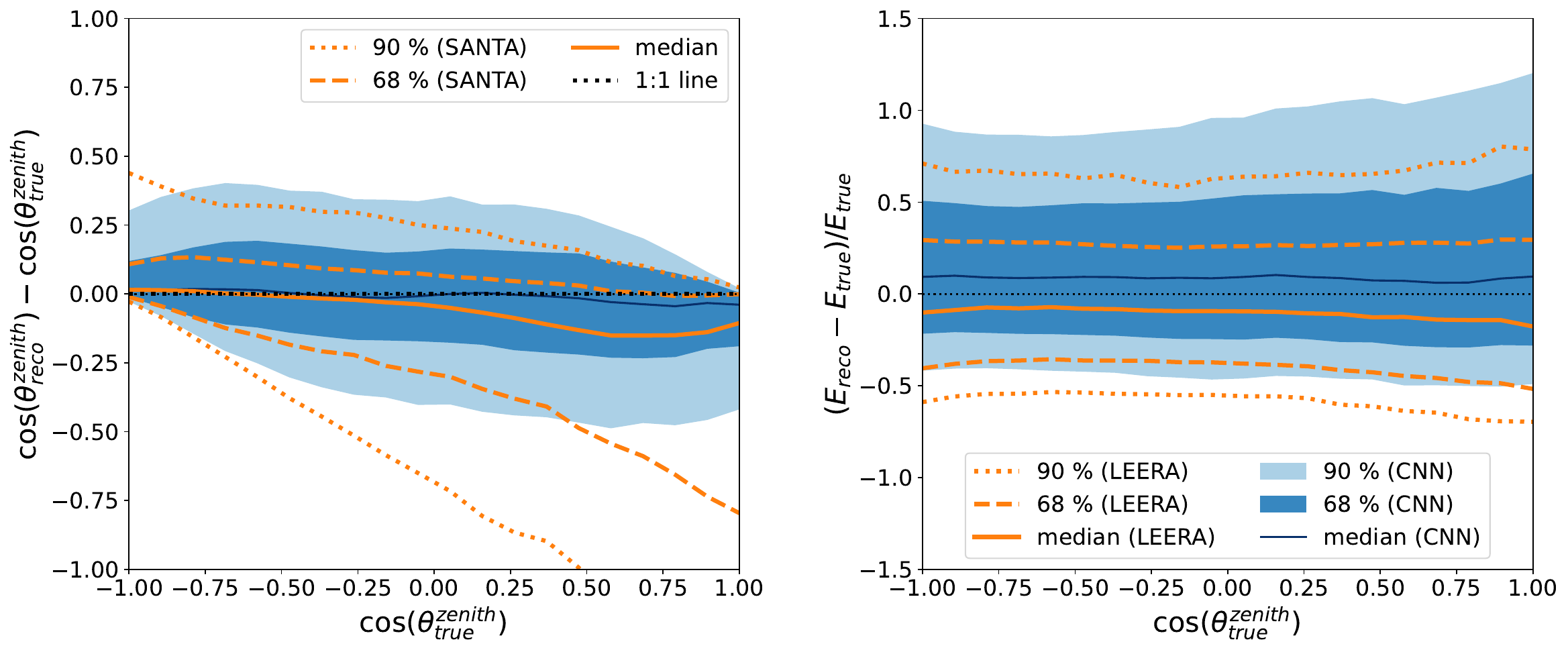}
\captionof{figure}{\label{fig:santa_flercnn_numu} Distribution of $\nu_{\mu}$ CC events of reconstructed neutrino zenith angle on the left and neutrino energy on the right plotted against true neutrino energy on the top and true neutrino arrival direction ($\cos(\theta^{zenith})$) on the bottom, with the median as solid curves, 68\% and 90\% quintiles as dark shaded  and light shaded for CNN, or dashed and dotted orange curves for SANTA/LEERA.} 
\end{figure}

\subsection{Muon Background Classifier}
To classify the atmospheric muon background in the neutrino sample, a mixed sample of $\nu_\mu$ CC, $\nu_e$ CC, and atmospheric muon events was prepared. 
The nominal muon simulation was split into a 10\% subsample used for training and the remaining 90\% was retained for analysis. In the processing of training datasets, the cut of BDT$_\mu>$0.95 was applied to the neutrino training dataset but not to the muons to retain enough statistics for training. 
Instead, a quality cut requiring $N_{\rm{pulse}}$ to be no less than four was applied to both the neutrino and muon training datasets. 
After processing, the training set for atmospheric muon classification had a ratio of 40:40:20 among atmospheric muon:$\nu_\mu$:$\nu_e$, where both CC and NC events were included in the neutrino portion. The total sample size was approximately 7 million events.

\begin{figure}[!tbh]
\centering
\includegraphics[width=\linewidth]{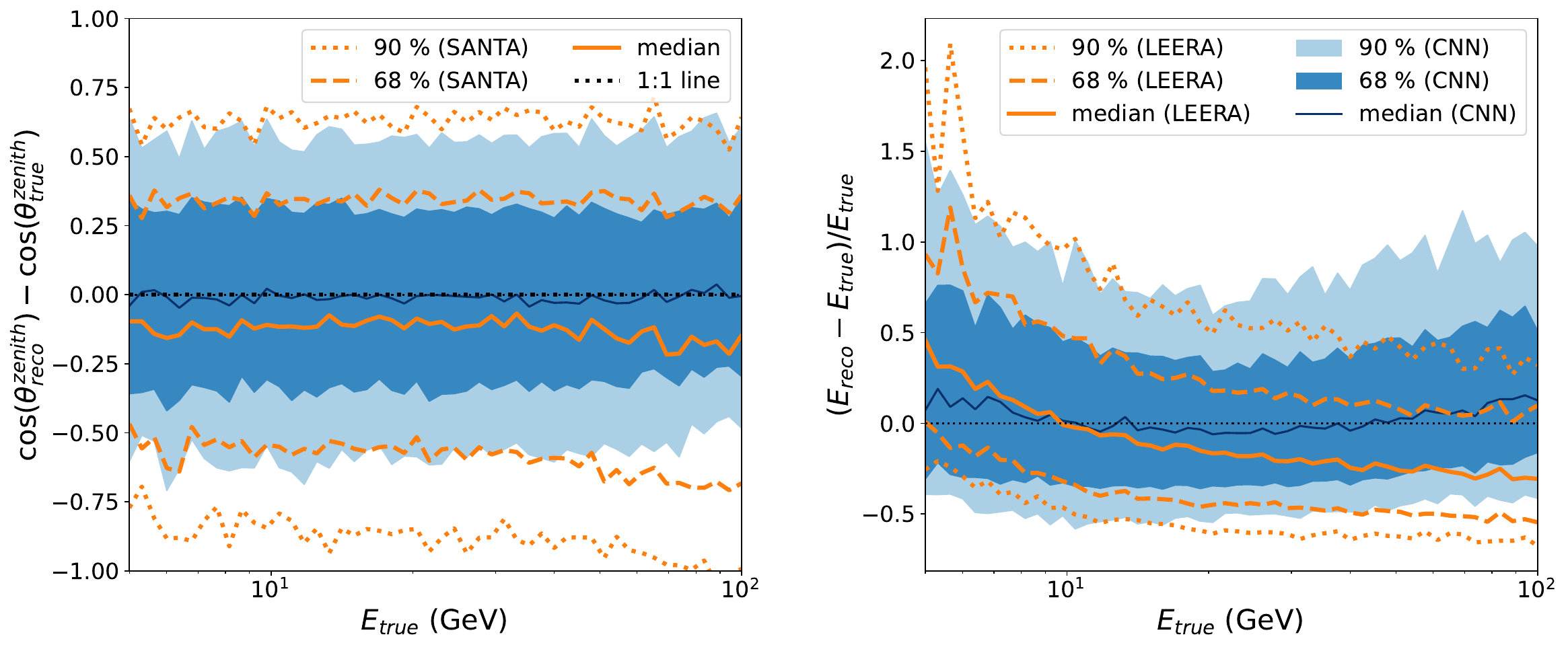}
\includegraphics[width=\linewidth]{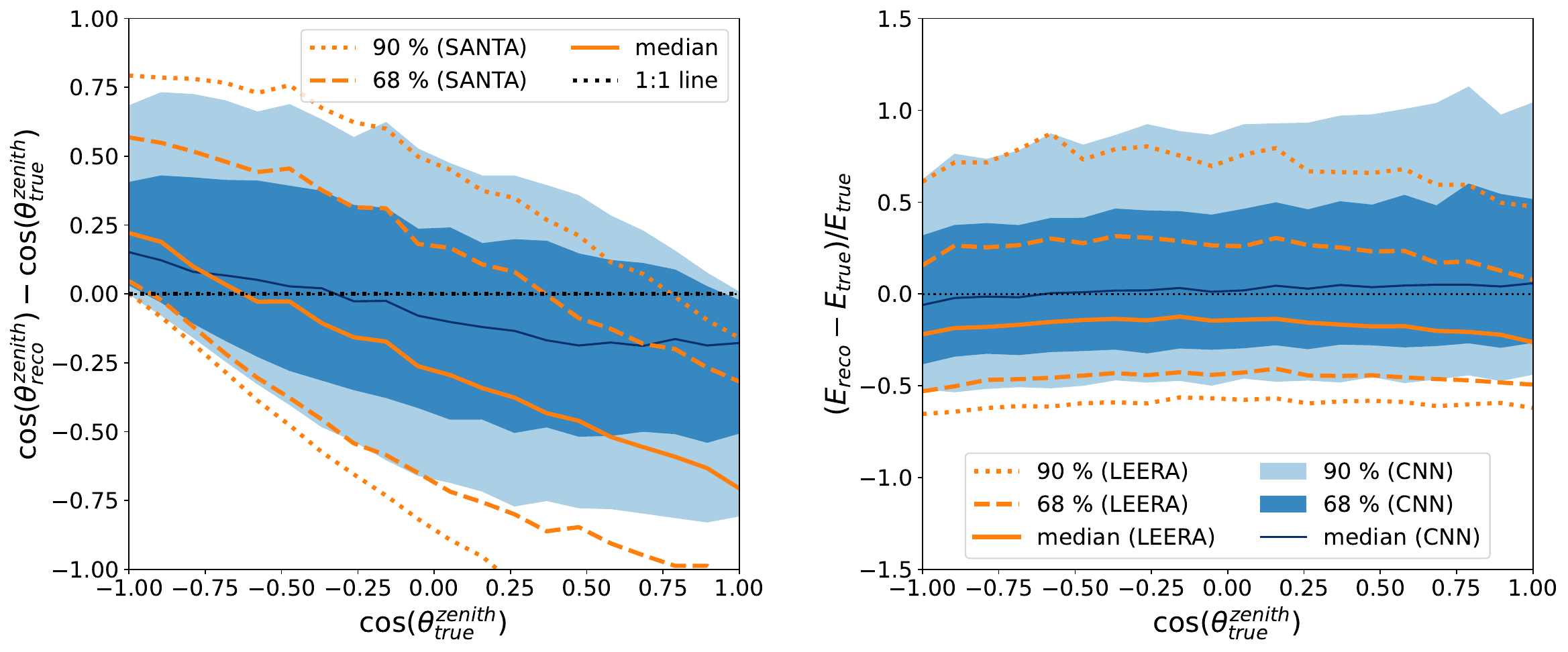}
\captionof{figure}{\label{fig:santa_flercnn_nue} Distribution of $\nu_{e}$ CC events of reconstructed neutrino zenith angle on the left and neutrino energy on the right plotted against true neutrino energy on the top and true neutrino arrival direction ($\cos(\theta_{zenith})$) on the bottom, with the median as solid curves, 68\% and 90\% quantiles as dark shaded  and light shaded for CNN, or dashed and dotted orange curves for SANTA/LEERA.} 
\end{figure} 
\section{Performance}\label{sec:perform}
The performance is evaluated on the MC dataset used in the oscillation analysis described in~\cite{paper_flercnn} and listed in Table II in~\cite{paper_verification}, with atmospheric flux, neutrino interactions, and oscillation weights applied. 
The previous IceCube neutrino oscillation result~\cite{paper_verification} used the SANTA tool for directional reconstruction and the LEERA tool for energy reconstruction, both described in IV.A of~\cite{PhysRevD.91.072004}. We compared the performance of our CNN routines with these tools using a common subset of events that passed the selections of both the current and previous oscillation analyses. 
To show the performance on neutrino energy and zenith angle (see Figures~\ref{fig:santa_flercnn_numu} and ~\ref{fig:santa_flercnn_nue}), the cuts on energy or zenith angle are not applied. The SANTA algorithm requires events to pass the selection criteria described in Section 2 of~\cite{le_reco_paper}, it is only applicable to a small set of high-quality events in the DeepCore detector. 
A more inclusive, likelihood-based algorithm called RETRO has been developed to provide more comprehensive information, allowing it to be applied to a larger set of events and yielding better performance~\cite{le_reco_paper}. However, the performance improvements achieved with RETRO come at the cost of significantly slower processing times compared to SANTA/LEERA. 
The CNNs presented in this paper, which are designed and optimized for the DeepCore region, address the slow processing time issue but offer comparable performance. A comparison of the runtime for SANTA, LEERA, RETRO and the CNNs discussed in this paper, is given in Table~\ref{tab:recotimes}. 
A detailed comparison between CNN and RETRO reconstructions can be found in Appendix~\ref{app:cnnvsretro}. 

\subsection{Comparison to Reconstruction used in the Prior DeepCore Oscillation Result}

\begin{figure}[!bth]
\begin{minipage}[!t]{0.45\linewidth}
\includegraphics[width=\linewidth]{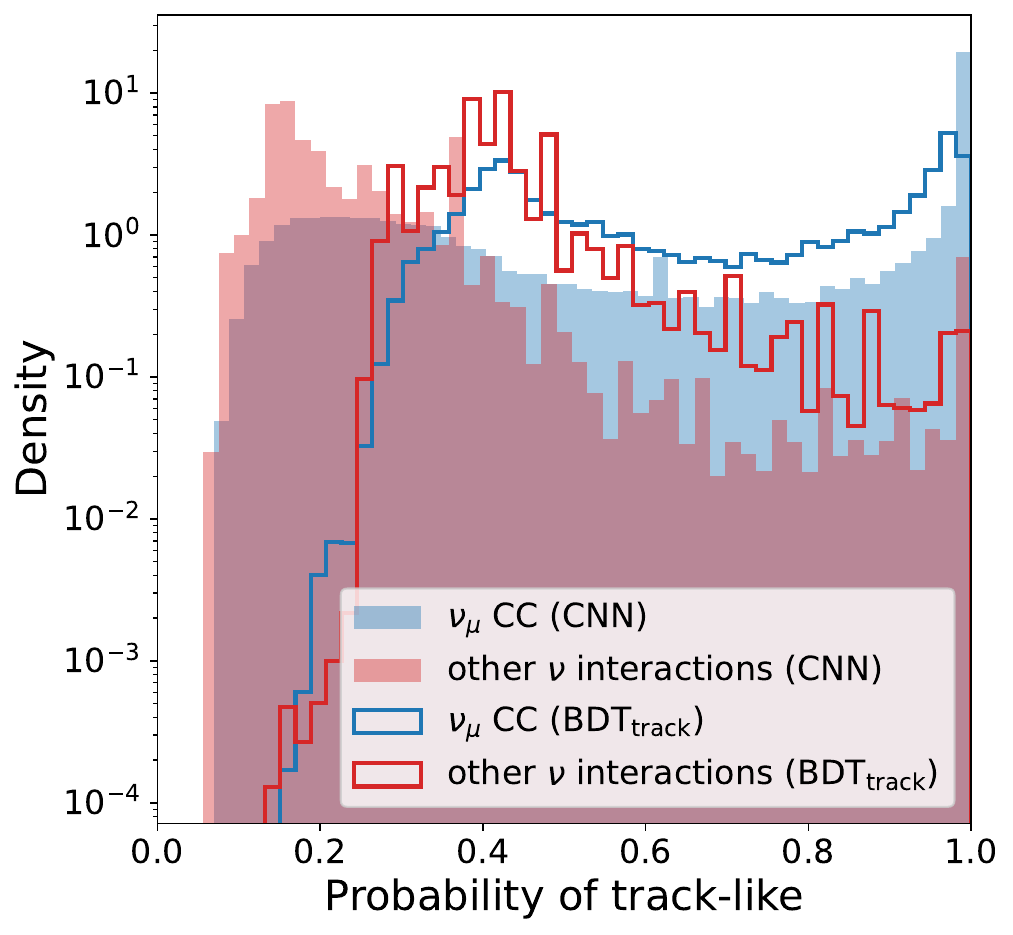}
\caption{Distributions of events classified by CNN as track-like or cascade-like, where most track-like corresponds to 1, with $\nu_\mu$ CC events in blue and other types of neutrino interactions in red, where filled histograms show CNN reconstruction and stepped histograms show BDT$_{\rm{track}}$.}\label{fig:PID}
\end{minipage}
  \hfill
\begin{minipage}[!t]{0.45\linewidth}
\includegraphics[width=\linewidth]{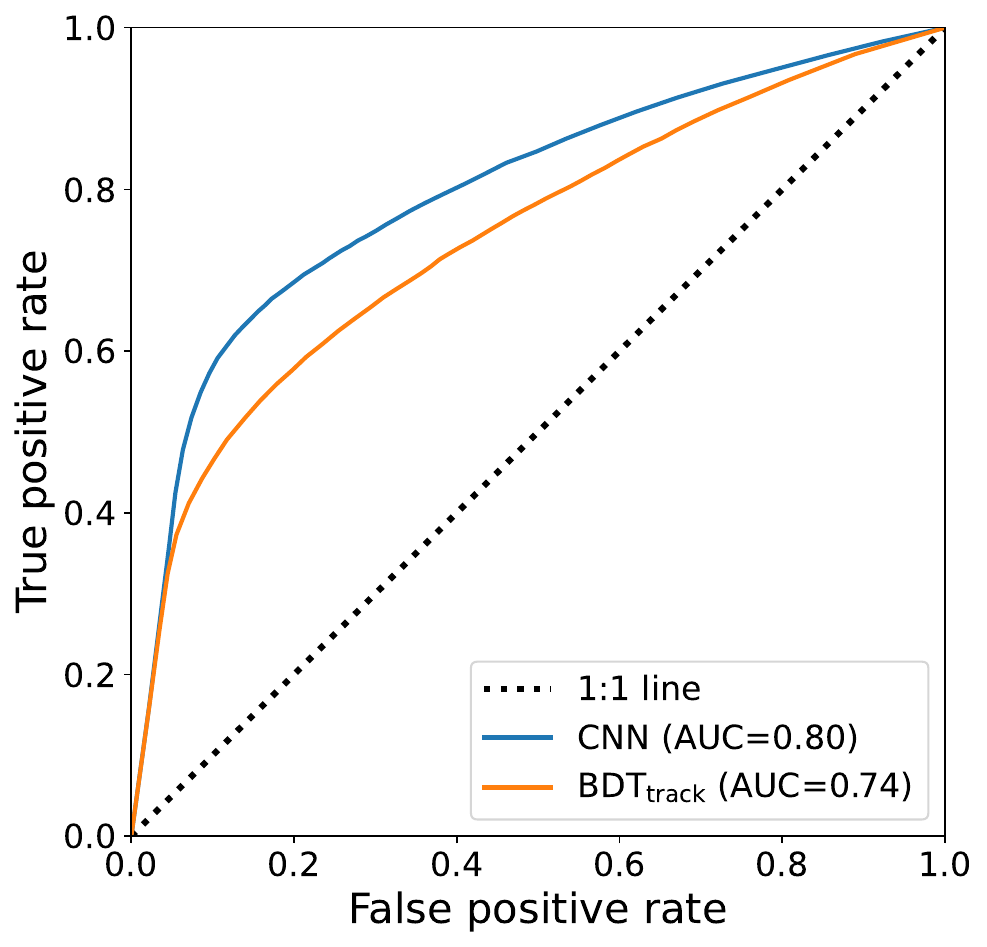}
\caption{ROC curves of BDT$_{\rm{track}}$ classifiers tested on a sample of events. CNN reconstruction is in blue and BDT$_{\rm{track}}$ is in orange.  \\ \\ \\ \ \ \ \                                                                 
   }\label{fig:roc_pid}
\end{minipage}

\end{figure}

Figure~\ref{fig:santa_flercnn_numu} and ~\ref{fig:santa_flercnn_nue} show the comparison of the reconstruction performance of the CNNs to the SANTA and LEERA methods used in the prior oscillation result~\cite{paper_verification} for both track-like and cascade-like events.
The median curves of the distributions of the CNN-reconstruction are flat and close to zero at different true neutrino energies (in the top panels) or true neutrino arrival directions (in the bottom panels). 
The comparisons of the particle identification and atmospheric muon classifiers are made on the common events that passed the selections of both analyses (excluding the cuts on neutrino energy, zenith angle, and CNN-reconstructed variables). In the last result (see Section IV.B of ~\cite{paper_verification}), a BDT (BDT$_{\rm{track}}$) was trained for PID classification (between track-like vs.\ cascade-like neutrino events). In Figure~\ref{fig:PID}, a comparison is shown between BDT$_{\rm{track}}$ and CNN on true $\nu_\mu$ CC (track-like) events and other types of neutrino interactions. The corresponding Receiver Operating Characteristic (ROC) curves of the two classifiers and the Area Under Curve (AUC) values can be found in Figure~\ref{fig:roc_pid}. 

True positive refers to that BDT$_{\rm{track}}$ correctly identifies as belonging to the positive class (track-like) while false positives are cascade events that are incorrectly identified as track-like events.

\begin{figure}[!b]
  \begin{minipage}[t]{0.48\textwidth}
    \centering
    \adjustbox{valign=t}{\includegraphics[width=\linewidth]{{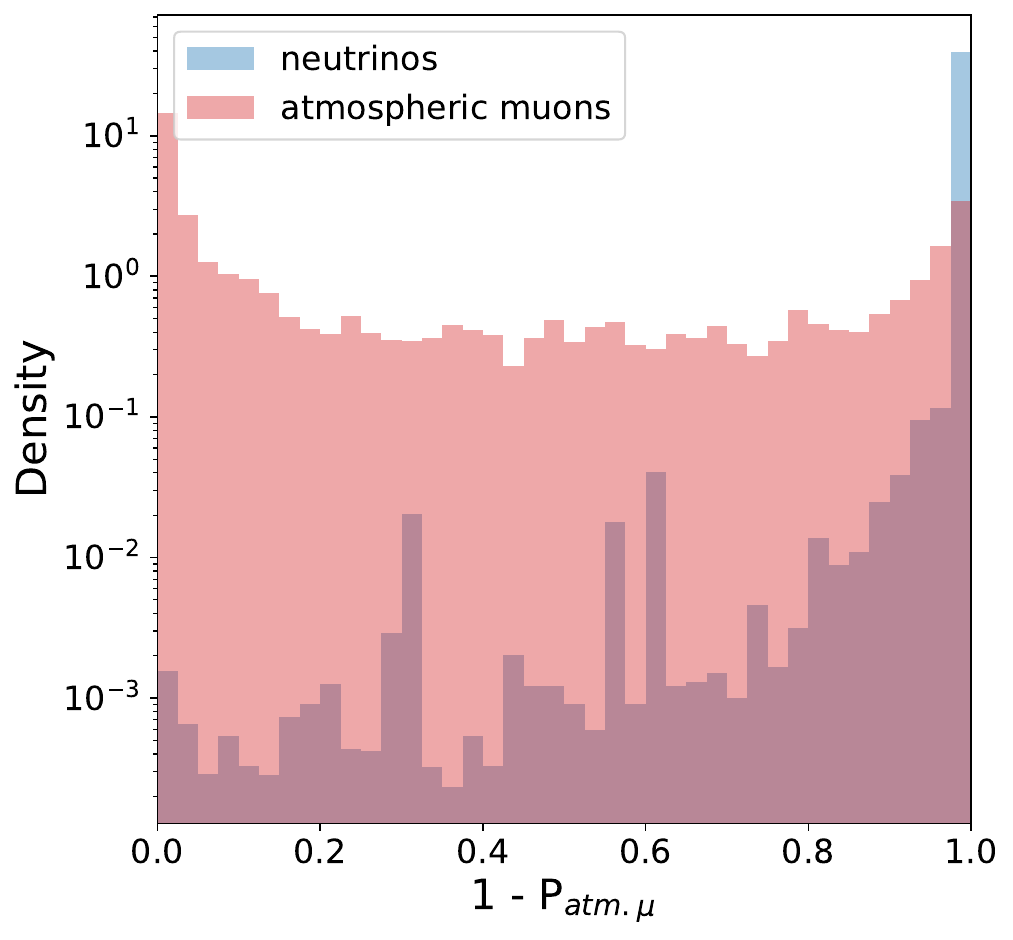}}}
    \caption{Distribution of events classified by CNN as neutrino-like, where most neutrino-like corresponds to 1, with $\nu_\mu$ CC events in blue and simulated atmospheric muons in red.}\label{fig:muonID}
  \end{minipage}
  \hfill
  \begin{minipage}[t]{0.47\textwidth}
    \adjustbox{valign=t}{\includegraphics[width=\linewidth]{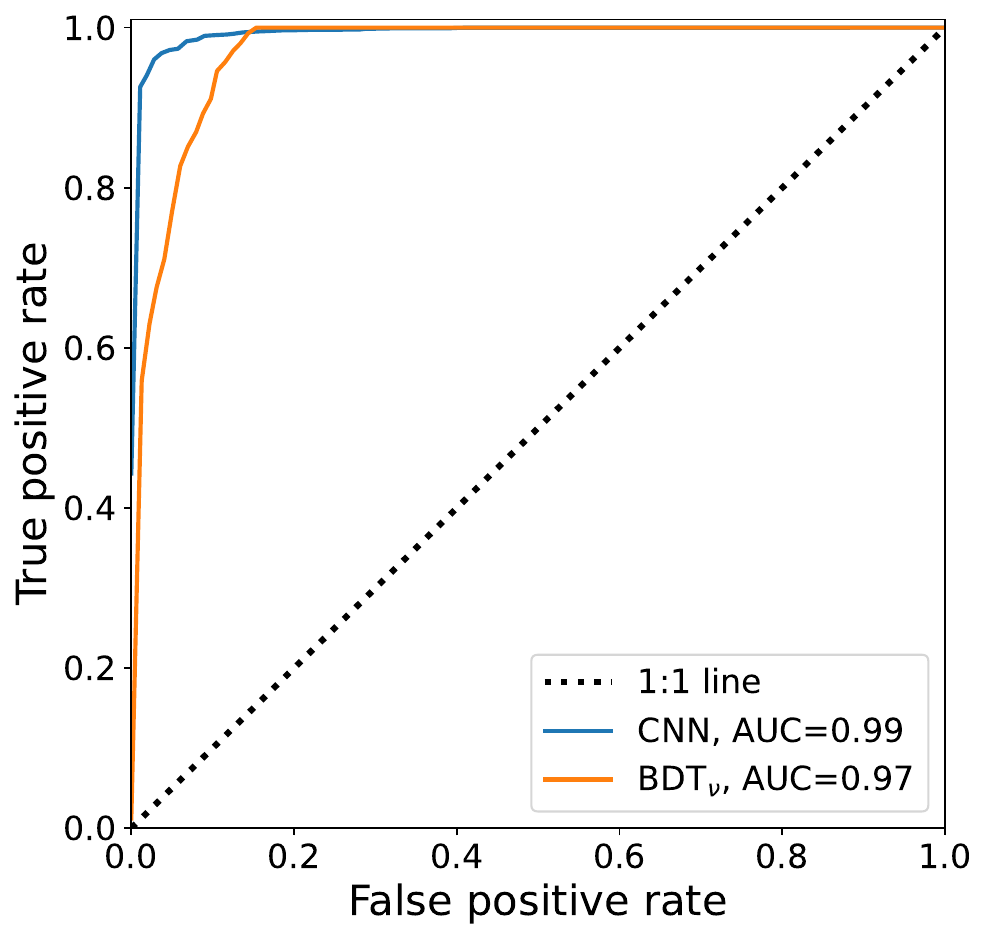}}
   \caption{ROC curves of atmospheric muon classifiers tested on a sample of events. CNN reconstruction is in blue and BDT$_\mu$ is in orange. \\ \\ \ \ \ \ }\label{fig:roc_muon}
  \end{minipage}
\end{figure}

The AUC values suggest that the CNN-reconstructed PID classifier can better distinguish $\nu_\mu$ CC from the other neutrino interactions on the common events. 
In Figure~\ref{fig:muonID}, the performance of the atmospheric muon classifier is shown as distributions of $1-P_{atm.\ \mu}$ with the true neutrino events peaking towards 1 (most neutrino-like). The neutrino/atmospheric muon classifier used in the previous analysis was also applied to the CNN training set, with a threshold of BDT$_\mu$ > 0.95. More details about the training and performance of the BDT$_\mu$ used in the last analysis can be found in Figures 7 and 16 in~\cite{paper_verification}. The discussion here is focused on the comparison of the common events.

Figure~\ref{fig:muonID} shows the common events that passed the selections of both analyses (without applying cuts on neutrino energy, zenith angle, or CNN-reconstructed variables). The ROC curves and AUC values for these selected common events are presented in Figure~\ref{fig:roc_muon}, where the AUC values indicate that both muon classifiers are effective at removing atmospheric muons. Since the CNN was trained with the BDT$_\mu$ cut applied, it further enhances the rejection of atmospheric muons that may have passed the initial BDT$_\mu$ threshold.

Besides the variables shown above, at the analysis level there are some variables, such as interaction vertex position ($x$, $y$, $z$) that are reconstructed by the CNN and employed as analysis cuts in~\cite{paper_flercnn} but they do not have corresponding predecessors in~\cite{paper_verification}. For these variables, we compare their performance with that of the likelihood-based reconstruction technique, RETRO~\cite{le_reco_paper}, which is a multipurpose likelihood-based reconstruction. In addition to reconstructing the interaction vertex position of neutrinos, RETRO offers better resolution than SANTA/LEERA for neutrino direction and energy estimation, at the expense of larger computational cost. Detailed comparisons can be found in Appendix~\ref{app:cnnvsretro}.

Besides the improvement in reconstruction performance in the selected analysis samples, the CNNs show an advantage in processing time since they can run on graphics processing units (GPU) (see Table~\ref {tab:recotimes}).
The average time per event was measured by running the reconstruction algorithms on a large sample of MC events. For SANTA and RETRO, the runtime depends on the true neutrino energy; therefore, we report the overall mean reconstruction time averaged over the full true neutrino energy spectrum. This is then compared to the average runtime of running the 5 CNNs on a similarly large sample of events within the same energy range.
The GPU used for evaluation was a Tesla V100S with a central processing unit (CPU) of Intel(R) Xeon(R) Platinum 8260 CPU @ 2.40 GHz. The same CPU was used to evaluate the inference speed of the 5 CNNs. The precision of the CNN parameters is float32. 

\begin{table}[!tb]
\centering{
\caption{\label{tab:recotimes}Runtime of applying the different reconstructions to the MC sample, where the last two columns are scaled from the first one to get a sensible estimate on a large statistics of production. The evaluation of the 5 CNNs is run sequentially.}}
\begin{tabular}{|c|c|c|c|} 
 \hline
 \textbf{Reconstruction} & \textbf{\begin{tabular}{@{}c@{}}Avg.\ time \\per event (s)\end{tabular}} & \textbf{\begin{tabular}{@{}c@{}}Events per day \\ (single core)\end{tabular}} & \textbf{\begin{tabular}{@{}c@{}}Time for 10$^8$ events \\ (1000 CPU cores or GPU nodes)\end{tabular}} \\[0.5ex] 
 \hline
5 CNNs on GPU & 0.0011  & 8$\times10^7$ & 2 min \\
 \hline
5 CNNs on CPU  & 0.29 & 3$\times10^5$ & 8 hours \\
\hline
SANTA and LEERA & 0.16~\cite{le_reco_paper}  & 5$\times10^5$  & 4 hours \\
\hline
RETRO & 40~\cite{le_reco_paper} & 2$\times10^3$ & 46 days \\
\hline
\end{tabular}
\end{table}

\section{Summary}\label{sec:summary}

The outcomes of this study show improvements in neutrino energy, zenith angle, and PID score, demonstrating better precision compared to previous reconstruction techniques for events that passed the final-level analysis cuts. This leads to a higher event rate after sample selection based on CNN-reconstructed variables. 
These CNN-based reconstructions are also used for binning the analysis sample. Furthermore, the CNNs developed in this study have been applied to interaction vertex reconstruction and atmospheric muon classification, which are crucial for selecting high-quality neutrino events in the most recent atmospheric neutrino oscillation analyses~\cite{paper_flercnn}, contributing to one of the most precise measurements of the $\nu_\mu$ disappearance parameters. Notably, this study has shown that the CNN approach achieves comparable reconstruction performance to RETRO, with the added benefit of a faster runtime, making it advantageous for processing large datasets.

Ongoing efforts include adapting low-energy CNNs for reconstructing low-energy events, with a focus on azimuth and estimating reconstruction uncertainty. These developments will be valuable for future IceCube analyses in both particle physics and astrophysics. The IceCube collaboration continues to prioritize improving reconstruction resolution and runtime through the development of different algorithms. At TeV energies, a CNN takes the whole detector for cascade reconstruction into account~\cite{Abbasi_2021}. Focusing on sub-100 GeV energies, we developed a CNN that uses DeepCore and surrounding IceCube strings, indicated by the orange hexagon in Figure~\ref{fig:detector}. 
Additionally, efforts are underway to develop an atmospheric neutrino sample using graph neural networks~\cite{gnn}.

\appendix

\section{Comparison to RETRO Reconstruction}\label{app:cnnvsretro}

\begin{figure}[!htb]
\includegraphics[width=\columnwidth]{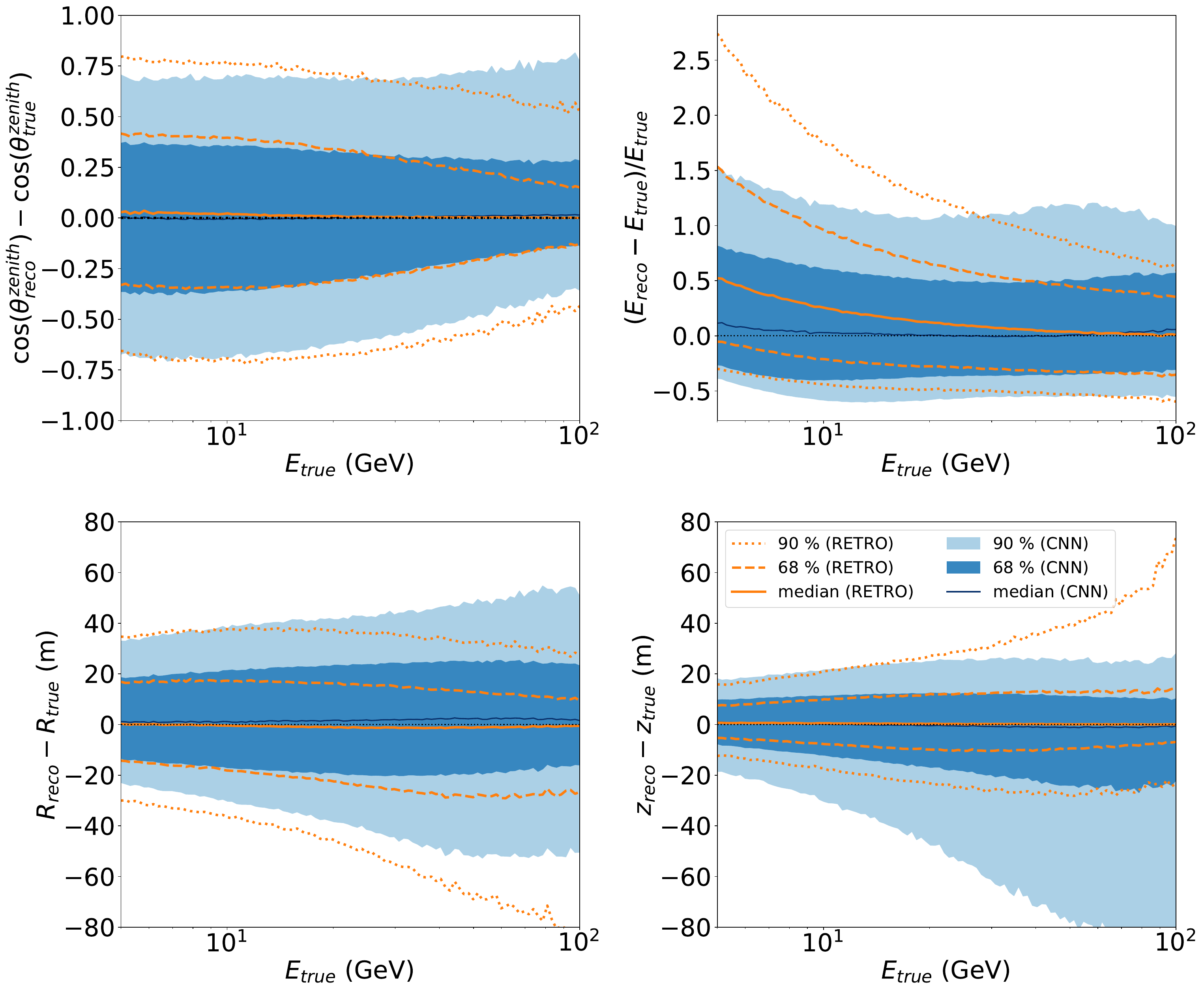}
\caption{\label{fig:retro_numu} Distribution of $\nu_{\mu}$ CC events of CNN reconstructed neutrino energy, zenith angle, radius (R), and $z$ position of neutrino interaction vertex against true neutrino energy. The medians are as solid curves, 68\% of events are as dark shaded, and 90\% of events are as light shaded. CNN reconstructions are in blue and RETRO reconstructions are in orange.} 
\end{figure}

\begin{figure}[!htb]
\includegraphics[width=\columnwidth]{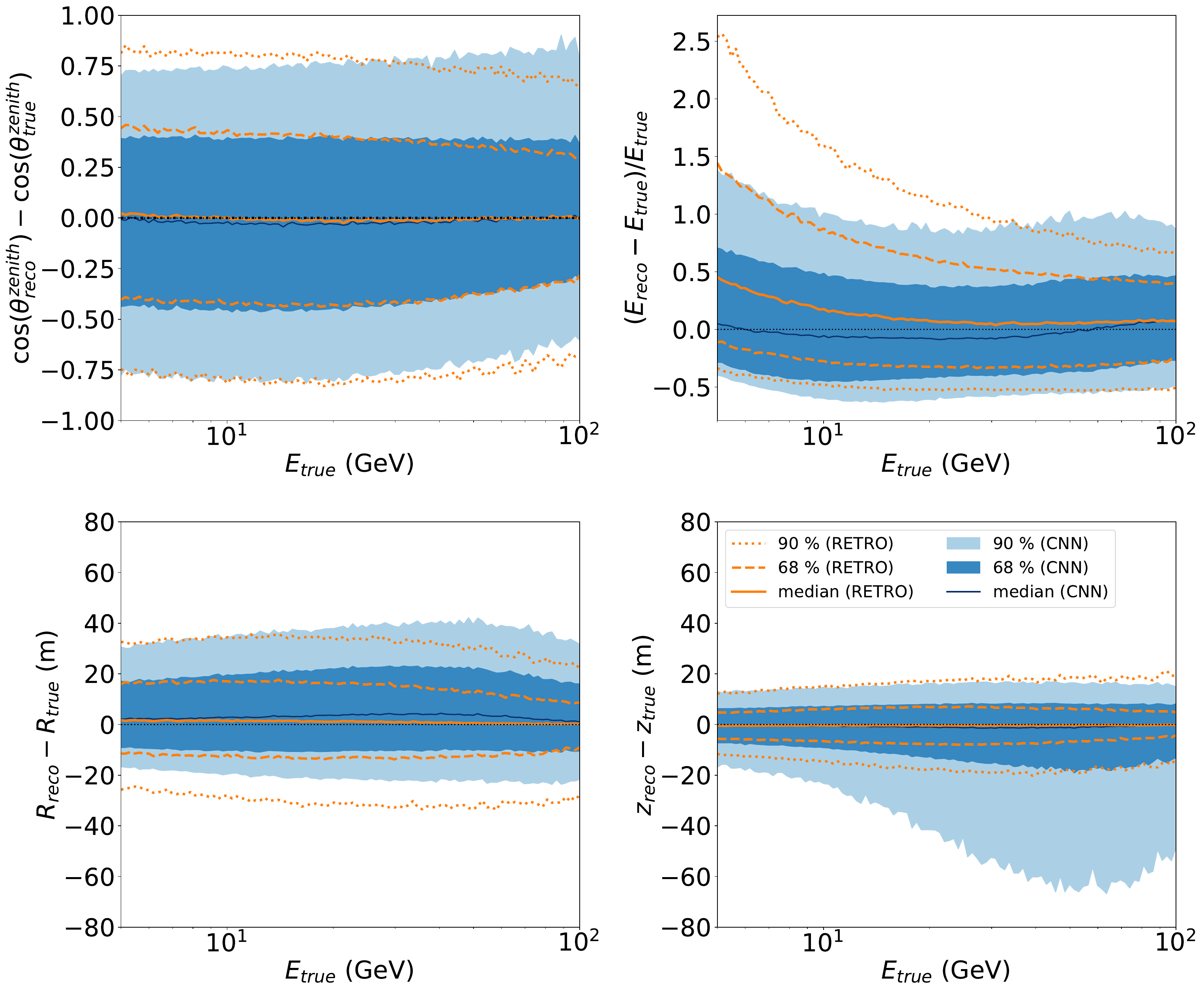}
\caption{\label{fig:nueCC_res} Distribution of $\nu_{e}$ CC events of CNN reconstructed neutrino energy, zenith angle, radius (R), and $z$ position of neutrino interaction vertex against true neutrino energy with the median as solid curve, 68\% as dark shaded, and 90\% as light shaded comparing to those of RETRO reconstructed in orange.} 
\end{figure}

Figure~\ref{fig:retro_numu} shows the comparisons of CNN and RETRO on a subset of overlapping events that pass all the cuts until the analysis-level, and the likelihood fitting quality cut of RETRO with flux, neutrino interaction cross-section, and oscillation weights applied.
The CNN reconstructs the interaction position in ($x$, $y$, $z$) coordinates, but the analysis cut is applied on depth ($z$) and horizontal radius relative to the position of the IceCube string-36 ($\rho_{36}$). Therefore, we show the performance of $z$, $\rho_{36}$, $\cos(\theta_{\rm{zenith}})$, and energy along the true neutrino energy as in Figure~\ref{fig:retro_numu}. In the $\nu_{\mu}$ disappearance analysis~\cite{paper_flercnn}, the signal events are $\nu_\mu$ CC. 
To optimize reconstruction performance for these signal events, the CNNs for neutrino energy, zenith angle, and interaction vertex were trained and tested on $\nu_\mu$ CC events. Similar evaluation plots for the CNN reconstruction of cascade-like events, using $\nu_{e}$ CC as an example, can be found in Figure~\ref{fig:nueCC_res}, which also shows reasonable performance.
Among all the performance plots, the asymmetric tails in the vertex $z$ distribution arise because good-quality events, which impact the CNN training more effectively, typically have vertices inside the DeepCore region. In contrast, higher-energy events are more likely to originate outside the lower part of DeepCore, leading the CNN to favor deeper $z$ predictions.

\acknowledgments
The IceCube collaboration acknowledges the significant contributions to this manuscript from Jessie Micallef and Shiqi Yu.
The authors gratefully acknowledge the support from the following agencies and institutions:
USA {\textendash} U.S. National Science Foundation-Office of Polar Programs,
U.S. National Science Foundation-Physics Division,
U.S. National Science Foundation-EPSCoR,
U.S. National Science Foundation-Office of Advanced Cyberinfrastructure, 
U.S. National Science Foundation AI Institute for Artificial Intelligence and Fundamental Interactions,
Wisconsin Alumni Research Foundation,
Center for High Throughput Computing (CHTC) at the University of Wisconsin{\textendash}Madison,
Open Science Grid (OSG),
Partnership to Advance Throughput Computing (PATh),
Advanced Cyberinfrastructure Coordination Ecosystem: Services {\&} Support (ACCESS),
Frontera and Ranch computing project at the Texas Advanced Computing Center,
U.S. Department of Energy-National Energy Research Scientific Computing Center,
Particle astrophysics research computing center at the University of Maryland,
Institute for Cyber-Enabled Research at Michigan State University,
Astroparticle physics computational facility at Marquette University,
NVIDIA Corporation,
and Google Cloud Platform;
Belgium {\textendash} Funds for Scientific Research (FRS-FNRS and FWO),
FWO Odysseus and Big Science programmes,
and Belgian Federal Science Policy Office (Belspo);
Germany {\textendash} Bundesministerium f{\"u}r Bildung und Forschung (BMBF),
Deutsche Forschungsgemeinschaft (DFG),
Helmholtz Alliance for Astroparticle Physics (HAP),
Initiative and Networking Fund of the Helmholtz Association,
Deutsches Elektronen Synchrotron (DESY),
and High Performance Computing cluster of the RWTH Aachen;
Sweden {\textendash} Swedish Research Council,
Swedish Polar Research Secretariat,
Swedish National Infrastructure for Computing (SNIC),
and Knut and Alice Wallenberg Foundation;
European Union {\textendash} EGI Advanced Computing for research;
Australia {\textendash} Australian Research Council;
Canada {\textendash} Natural Sciences and Engineering Research Council of Canada,
Calcul Qu{\'e}bec, Compute Ontario, Canada Foundation for Innovation, WestGrid, and Digital Research Alliance of Canada;
Denmark {\textendash} Villum Fonden, Carlsberg Foundation, and European Commission;
New Zealand {\textendash} Marsden Fund;
Japan {\textendash} Japan Society for Promotion of Science (JSPS)
and Institute for Global Prominent Research (IGPR) of Chiba University;
Korea {\textendash} National Research Foundation of Korea (NRF);
Switzerland {\textendash} Swiss National Science Foundation (SNSF).

\bibliographystyle{JHEP}
\bibliography{biblio.bib}

\end{document}